\documentclass[article, twocolumn,
   aps, pra,
  amsmath,amssymb,
  longbibliography,
  ]{revtex4-1}
\usepackage{graphicx,color}
\usepackage{amsmath}
\usepackage{natbib}
\usepackage{epsfig}
\begin{document}

\title{Excess entropy of strongly coupled Yukawa fluids}

\author{S. A. Khrapak}\email{Sergey.Khrapak@gmx.de}
\affiliation{Joint Institute for High Temperatures, Russian Academy of Sciences, 125412 Moscow, Russia}

\begin{abstract}
The excess entropy of strongly coupled Yukawa fluids is discussed from several perspectives. First, it is demonstrated that a vibrational paradigm of atomic dynamics in dense fluids can be used to obtain simple and accurate estimate of the entropy without any adjustable parameters. Second, it is explained why a quasi-universal value of the excess entropy of simple fluids at the freezing point should be expected, and it is demonstrated  that a remaining very weak dependence of the freezing point entropy on the screening parameter in the Yukawa fluid can be described by a simple linear function. Third, a scaling of the excess entropy with the freezing temperature is examined, a modified form of the Rosenfeld-Tarazona scaling is put forward, and some consequences are briefly discussed. Fourth, the location of the Frenkel line on the phase diagram of Yukawa systems is discussed in terms of the excess entropy and compared with some predictions made in the literature. Fifth, the excess entropy scaling of the transport coefficients (self-diffusion, viscosity, and thermal conductivity) is re-examined using the contemporary data sets for the transport properties of Yukawa fluids. The results can be of particular interest in the context of complex (dusty) plasmas, colloidal suspensions, electrolytes, and other related systems with soft pairwise interactions.               
\end{abstract}

\date{\today}

\maketitle

\section{Introduction}

Entropy is a central concept in thermodynamics and statistical mechanics. In statistical physics it is defined as a measure of the number of possible microscopic states (microstates) of a system in thermodynamic equilibrium, at given macroscopic thermodynamic properties~\cite{LandauStatPhys,SatorStatPhys}. This connection between macroscopic and microscopic behavior is vital for understanding structural and dynamical properties of gases, liquids, and solids. While in gases and solids the entropy can be readily evaluated, this is not the case in liquids. An accurate equation of state is required to evaluate the entropy by conventional thermodynamic integration. An accurate equation of state is not always available and other methods can be sometimes useful.

The main purpose of this paper is to apply a vibrational paradigm of atomic dynamics in dense liquids to the strongly coupled Yukawa fluid. The vibrational picture~\cite{KhrapakPhysRep2024} has been rather successful in describing certain transport properties of various simple fluids, such as Stokes-Einstein relation between the self-diffusion and viscosity coefficients~\cite{KhrapakPRE10_2021,KhrapakMolecules12_2021,KhrapakJCP04_2024}, as well as the thermal conductivity coefficient~\cite{KhrapakPRE01_2021,KhrapakPoP01_2021,KhrapakPoP08_2021,KhrapakPPR2023,KhrapakPRE12_2023}.   
Within this paradigm, the solid-like oscillations of atoms around their temporary equilibrium positions dominate the dynamical picture. The temporary equilibrium positions of atoms do not form any regular structure
and are not fixed, unlike in solids. Instead, they are allowed to diffuse and this is why
liquids can flow. However, this diffusive motion is characterized by much longer time
scales compared to the period of solid-like oscillations. These long time scales are irrelevant from the point of view of the system entropy. Hence, the vibrational model has been demonstrated to describe relatively well the excess entropy of inverse-power-law fluids with $\propto r^{-n}$ repulsive interactions, including the extremely soft Coulomb ($n=1$) case, intermediately soft $n=6$ and $n=12$ cases, as well as the opposite hard-sphere interaction limit ($n=\infty$)~\cite{KhrapakJCP2021}. It is reasonable to expect that the same method would work when applied to the Yukawa fluid and we verify this conjecture below.       

The vibrational model is applied to the Yukawa fluid using the dispersion relations of collective modes from the quasi-localized charge approximation (QLCA)~\cite{Hubbard1969,GoldenPoP2000,DonkoJPCM2008} combined with a simple model of the radial distribution function~\cite{KhrapakPoP2016}. The results are compared with those from extensive molecular dynamics (MD) simulation of Refs.~\cite{FaroukiJCP1994,HamaguchiPRE1997} and good agreement is documented. Application of the Rosenfeld-Tarazona scaling to estimate the excess entropy is discussed and a new modified expression is put forward. This modified expression is used to discuss the location of the Frenkel line on the phase diagram of Yukawa fluids in terms of the excess entropy. Excess entropy scaling of transport coefficients proposed by Rosenfeld~\cite{RosenfeldPRA1977} is a very important corresponding states principle for the transport properties of fluids~\cite{DyreJCP2018}. We use this opportunity to demonstrate how it applies to Yukawa fluids using contemporary transport data from extensive numerical simulations. A complete and intrinsically consistent picture on how the phase state, dynamical, and transport properties of Yukawa fluids are inter-related via the excess entropy thus emerges.   

\section{Screened Coulomb (Yukawa) fluid}

The pairwise screened Coulomb repulsive potential (also referred to as Debye-H\"uckel or Yukawa potential) is usually used as a first approximation to describe interactions between charged particles in neutralizing media, such as plasmas, complex (dusty) plasmas, electrolytes, and colloidal suspensions~\cite{FortovUFN,FortovPR,
IvlevBook,KhrapakCPP2009,ChaudhuriSM2011,BeckersPoP2023,KonopkaPRL2000}. The potential is     
\begin{equation}\label{Yukawa}
\phi(r)=\frac{Q^2}{r}\exp\left(-\frac{r}{\lambda}\right),
\end{equation}
where $Q$ is electrical charge and $\lambda$ is the screening length. The properties of Yukawa systems are  described by the two dimensionless parameters. These are the Coulomb coupling parameter $\Gamma=Q^2/aT$ and the screening parameter $\kappa=a/\lambda$, where $T$ is the temperature in energy units, $a=(4\pi\rho/3)^{-1/3}$ is the Wigner-Seitz radius, and $\rho$ is the number density. The screening parameter $\kappa$ determines the softness of the Yukawa potential. It varies from the extremely soft and long-ranged Coulomb potential $\propto 1/r$ at $\kappa\rightarrow 0$, which corresponds to the one-component plasma limit, to the hard-sphere-like interaction limit at $\kappa\rightarrow \infty$. In the context of complex plasmas and colloidal suspensions the relatively ``soft'' regime, $\kappa\sim {\mathcal O}(1)$, is of particular interest. Most attention in the literature has been focused on this regime and we follow this tradition here.  

The properties of the phase diagram of Yukawa systems are relatively well understood. Since the Yukawa potential is purely repulsive,  there are no gas-liquid phase transition, gas-liquid coexistence, critical and gas-liquid-solid triple points. There is a fluid-solid phase transition at sufficiently strong coupling. The location of the freezing line $\Gamma_{\rm fr}(\kappa)$ for $\kappa<5$ is tabulated in Ref.~\cite{HamaguchiPRE1997} and simple practical fits are available~\cite{VaulinaJETP2000,VaulinaPRE2002}. There is also a crossover between gas-like and liquid-like dynamics, known as Frenkel line on the phase diagram~\cite{BrazhkinPRE2012,BrazhkinPRL2013,BellJCP2020,KhrapakJCP2022}. Frenkel line appears to be parallel to the freezing line and is located at $\Gamma/\Gamma_{\rm fr}\simeq 0.05$, as identified recently~\cite{HuangPRR2023,YuPRE2024}. We will elaborate on this topic below. At a very high coupling, the possibility of the glass transition has been predicted. The traditional mode coupling theory results from Refs.~\cite{YazdiPRE2014,CastelloMolecules2021} show that the glass transition line is also parallel to the freezing line and is predicted at $\Gamma/\Gamma_{\rm fr}\simeq 2-3$, depending on the chosen closure to the integral equation theory used to obtain structural information.

Transport phenomena and thermodynamics of three-dimensional Yukawa fluids have been investigated in dozens of publications, see e.g. Ref.~\cite{KhrapakPhysRep2024} for a recent overview. The main reference source for this paper are Refs.~\cite{FaroukiJCP1994,HamaguchiPRE1997} in what concerns the fluid-solid phase transition and thermodynamics of Yukawa fluids. Regarding the transport coefficients discussed later in the paper in the context of the excess entropy scaling, for the self-diffusion coefficient we use numerical data from Ref.~\cite{OhtaPoP2000}, for the shear viscosity coefficient -- data from Refs.~\cite{DonkoPRE2008,DaligaultPRE2014}, and for the thermal conductivity coefficient -- data from Refs.~\cite{DonkoPRE2004,ScheinerPRE2019}. Collective mode properties are derived from the quasi-crystalline approximation~\cite{Hubbard1969}, more familiar as the quasi-localized charge approximation (QLCA) in the plasma-related context~\cite{GoldenPoP2000,DonkoJPCM2008,KhrapakSciRep2017}.  

Dealing with thermodynamic quantities, we work only with their ``excess'' components, which are the actual quantity minus the ideal gas contribution at the same temperature and density~\cite{HansenBook}. Additionally, we work with dimensionless units. In particular, the reduced energy is $u=U/NT$, the reduced entropy is $s=S/Nk_{\rm B}$, and $c_{\rm V}=C_{\rm V}/Nk_{\rm B}$ is the specific heat at constant volume. 
 
To conclude this Section we briefly discuss another important aspect, related to excess entropy and Yukawa fluids. Excess entropy plays central role in the theory of ``isomorphs''~\cite{GnanJCP2009,DyreJPCB2014}. Isomorphs are defined as lines in the thermodynamic phase
diagram along which structure and dynamics in properly reduced units are invariant to a good approximation~\cite{DyreJPCB2014}. Constancy of excess entropy often serves as an identification of isomorphs. It can be rigorously proven that systems that have strong correlations between the equilibrium virial and potential energy fluctuations exhibit good isomorphs in their phase diagrams~\cite{GnanJCP2009,DyreJPCB2014,DyrePRE2013}. Such systems are usually referred to as ``Roskilde-simple'' or just ``R-simple'' systems~\cite{DyrePRE2013}. It has been demonstrated that the Yukawa system has rather strong correlations between the virial and potential energy and thus belongs to the class of R-simple systems~\cite{VeldhorstPoP2015,CastelloJCP2021}. Therefore, the excess entropy plays a particularly important role in Yukawa fluids.

\section{Vibrational model of the excess entropy at work}

Within the vibrational model, the reduced excess entropy of a liquid is obtained as an appropriate averaging over the frequencies of normal modes~\cite{KhrapakJCP2021}:
\begin{equation}\label{sex}
s_{\rm ex} = \frac{3}{2}-\frac{3}{2}\left\langle\ln \frac{m\Delta^2\omega^2}{2\pi T}\right\rangle,
\end{equation}
where $m$ is the mass, $\Delta=\rho^{-1/3}$ is the structure independent mean interparticle separation, and $\omega$ is the vibrational frequency.
Although this resembles an expression for the excess entropy of a crystalline solid, there is an important difference. The reduced liquid excess entropy contains an additional term $+1.0$ compared to the solid entropy, which is sometimes referred to as  ``communal entropy''~\cite{Hirschfelder1937}. In the vibrational model, this term naturally occurs from the assumption of single-cell occupancy, see Ref.~\cite{KhrapakJCP2021} for further detail.

The second term in Eq.~(\ref{sex}) can be rewritten as 
\begin{equation}\label{average1}
\left\langle\ln \frac{m\Delta^2\omega^2}{2\pi T}\right\rangle=\left\langle \ln\frac{\omega^2}{\omega_{\rm p}^2}\right\rangle + \ln\Gamma +\ln\frac{2}{(4\pi/3)^{1/3}},
\end{equation} 
where $\omega_{\rm p}=\sqrt{4\pi Q^2 \rho/m}$ is the plasma frequency.
Following the standard practice, we express averaging over frequencies as an integral over the wave-numbers~\cite{LandauStatPhys} 
\begin{equation}\label{average2}
\left\langle\ln \frac{\omega^2}{\omega_{\rm p}^2}\right\rangle=\frac{2}{9\pi}\int_0^{q_{\rm max}}q^2dq\left(\ln\frac{\omega_l^2}{\omega_{\rm p}^2}+2\ln\frac{\omega_t^2}{\omega_{\rm p}^2}\right).
\end{equation}  
Here $q = ka$ is the reduced wave number and $\omega_{l,t}(q)$ are the dispersion relations of the longitudinal and transverse modes, respectively. It is assumed that dense liquids support one longitudinal (compressional) and two transverse (shear) modes. The cut-off wave number $q_{\rm max}=(9\pi/2)^{1/3}\simeq 2.418$ ensures that the identity $\langle {\mathcal X} \rangle ={\mathcal X}$ is satisfied, where $\mathcal X$ is any quantity that does not depend on the wave number.  

\begin{figure*}
\includegraphics[width=17cm]{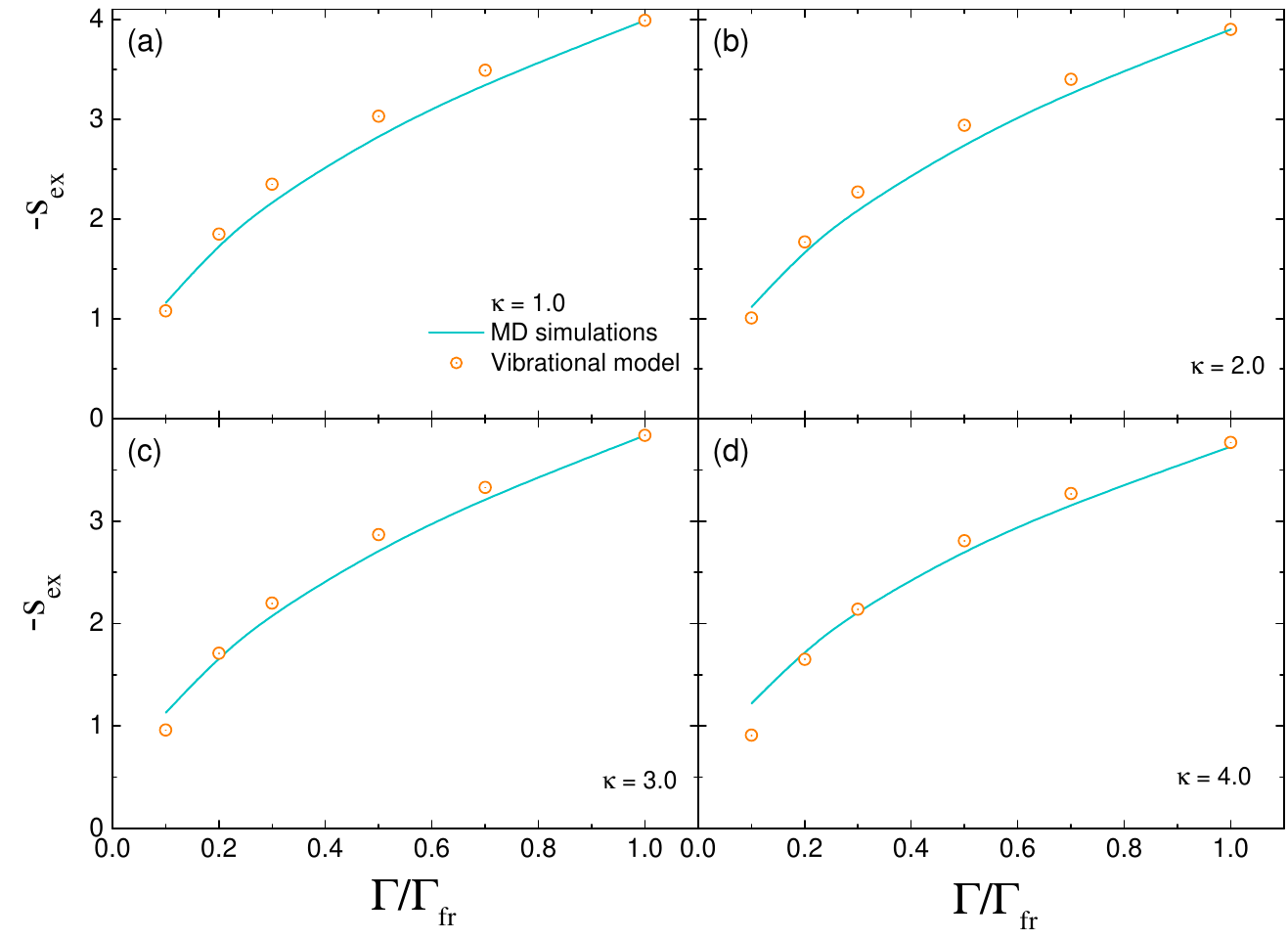}
\caption{(Color online) Negative reduced excess entropy of the Yukawa fluid as a function of reduced coupling parameter $\Gamma/\Gamma_{\rm fr}$. The results for four screening parameters are presented: $\kappa=1.0$ (a), $\kappa=2.0$ (b), $\kappa=3.0$ (c), $\kappa=4.0$ (d). The symbols correspond to the calculation using the vibrational model reported here. The solid curves are plotted using the results from MD simulations in Refs.~\cite{FaroukiJCP1994,HamaguchiPRE1997} (see Appendix for details).}
\label{FigEntropy}
\end{figure*}

We perform averaging using the dispersion relations derived within the framework of the QLCA model. Moreover, since the long-wavelength range of the dispersion relations is mostly involved (due to the cut-off), it is practical to employ the simplified version of the QLCA in which the radial distribution function (RDF) is modelled with a simple step function (excluded volume approximation). This result in simple analytical expressions, which demonstrate high accuracy at long wavelengths~\cite{KhrapakPoP2016,KhrapakAIPAdv2017,KhrapakIEEE2018}. These dispersion relations are  
\begin{equation}\label{L1}
\begin{aligned}
\frac{\omega_l^2}{\omega_{\rm p}^2}=\frac{q^2}{\Gamma}
& +e^{-R\kappa}\left[\left(1+R\kappa\right)\left(\frac{1}{3}-\frac{2\cos Rq}{R^2q^2}+\frac{2\sin Rq}{R^3q^3} \right) \right. \\ 
& \left. -\frac{\kappa^2}{\kappa^2+q^2}\left(\cos Rq+\frac{\kappa}{q}\sin Rq \right)\right],
\end{aligned}
\end{equation}  
and
\begin{equation}\label{T1}
\frac{\omega_t^2}{\omega_{\rm p}^2}=\frac{q^2}{3\Gamma}+e^{-R\kappa}\left(1+R\kappa\right)\left(\frac{1}{3}+\frac{\cos Rq}{R^2q^2}-\frac{\sin Rq}{R^3q^3} \right).
\end{equation}
Here $R$ is the excluded volume radius expressed in units of $a$. The excluded volume radius is chosen in such a way that the model RDF $g(r)=\theta(r-aR)$ yields accurate results via energy or pressure equations~\cite{KhrapakPoP2016}. A simple practical expression proposed in Ref.~\cite{KhrapakAIPAdv2017} is
\begin{equation}
R(\kappa)\simeq 1+ \frac{1}{\kappa}\ln\left[\frac{3\cosh(\kappa)}{\kappa^2}-\frac{3\sinh(\kappa)}{\kappa^3}\right].
\end{equation}       
Each of Eqs.~(\ref{L1}) and (\ref{T1}) contains two terms. The first is the kinetic contribution. The second is the potential energy contribution, associated with the interaction between the particles. In the strongly coupled regime the potential contribution dominates and kinetic terms are often omitted. We keep both contributions and this provides slightly better accuracy in estimating the excess entropy.         

Substituting Eqs.~(\ref{L1}) and (\ref{T1}) in formula (\ref{average2}) and performing numerical integration we obtain estimation of the excess entropy within the vibrational model. The results of our theoretical calculation along with the comparison with numerical results is presented in Fig.~\ref{FigEntropy}. The agreement is remarkable, especially taking into account the simplicity of the model and the absence of any adjustable parameters. 

It should be pointed out that the present approach neglects the so-called $q$-gap -- zero-frequency portion of the dispersion relation at low $q$ also known as the propagation gap in the dispersion relation of the transverse collective mode~\cite{HansenBook,OhtaPRL2000,MurilloPRL2000,GoreePRE2012,BrykPRE2014,
BolmatovPCL2015,TrachenkoRPP2015,YangPRL2017,KhrapakJCP2019,KryuchkovSciRep2019,KryuchkovJCP2021}. Given the good agreement between the present theory and the actual results for the excess entropy, there seems no need to further complicate the model by including the $q$-gap into consideration. This is a major difference from the so-called phonon theory of liquid thermodynamics~\cite{BolmatovSciRep2012,TrachenkoRPP2015}, where the $q$-gap and the disappearance of the transverse modes when approaching the gaseous state play a major role. On the other hand, the vibrational model discussed here is focused on dense liquid states, when the vibrational motion dominates, transverse modes are present, and the $q$-gap is relatively narrow. The limits of the applicability of the vibrational model are discussed further in this paper.

Given the fundamental significance of the excess entropy in R-simple fluids such as the Yukawa fluid, it would be very important in future to obtain accurate MD or MC simulation data for the excess entropy. The calculations based on the equation of state does not suffice, given the thermodynamic integration for the excess free energy that introduces uncontrolled errors. Such indirect evaluation is possible based on the fundamental thermodynamic Euler equation combined with separate
extractions of the excess internal energy, the excess pressure from the virial equation~\cite{AllenBook}  and the excess chemical potential via the Widom insertion method~\cite{WidomJCP1963,WidomJPC1982}. Note that isomorph theory allows the exact tracking of isentropic lines via numerical simulations, but does not allow the calculation of the excess entropy value along these lines. Nevertheless, it can be employed for rigorous cross-checking. 

\section{Excess entropy at freezing}

The excess entropy at freezing is a quasi-universal quantity for many simple fluids such as the inverse-power-law fluid~\cite{RosenfeldPRE2000}, Lennard-Jones fluid~\cite{KhrapakJPCL2022,KhrapakJCP2022_1}, as well as the Yukawa fluid considered here (see Fig.~\ref{FigEntropy}). In all cases $s_{\rm ex}\simeq -4$ applies (a somewhat smaller value for Yukawa fluids was reported in Ref.~\cite{JoyPoP2017}, the reason for this deviation is unclear). Let us look in more detail whether any systematic trend is present. 

\begin{figure}
\includegraphics[width=8cm]{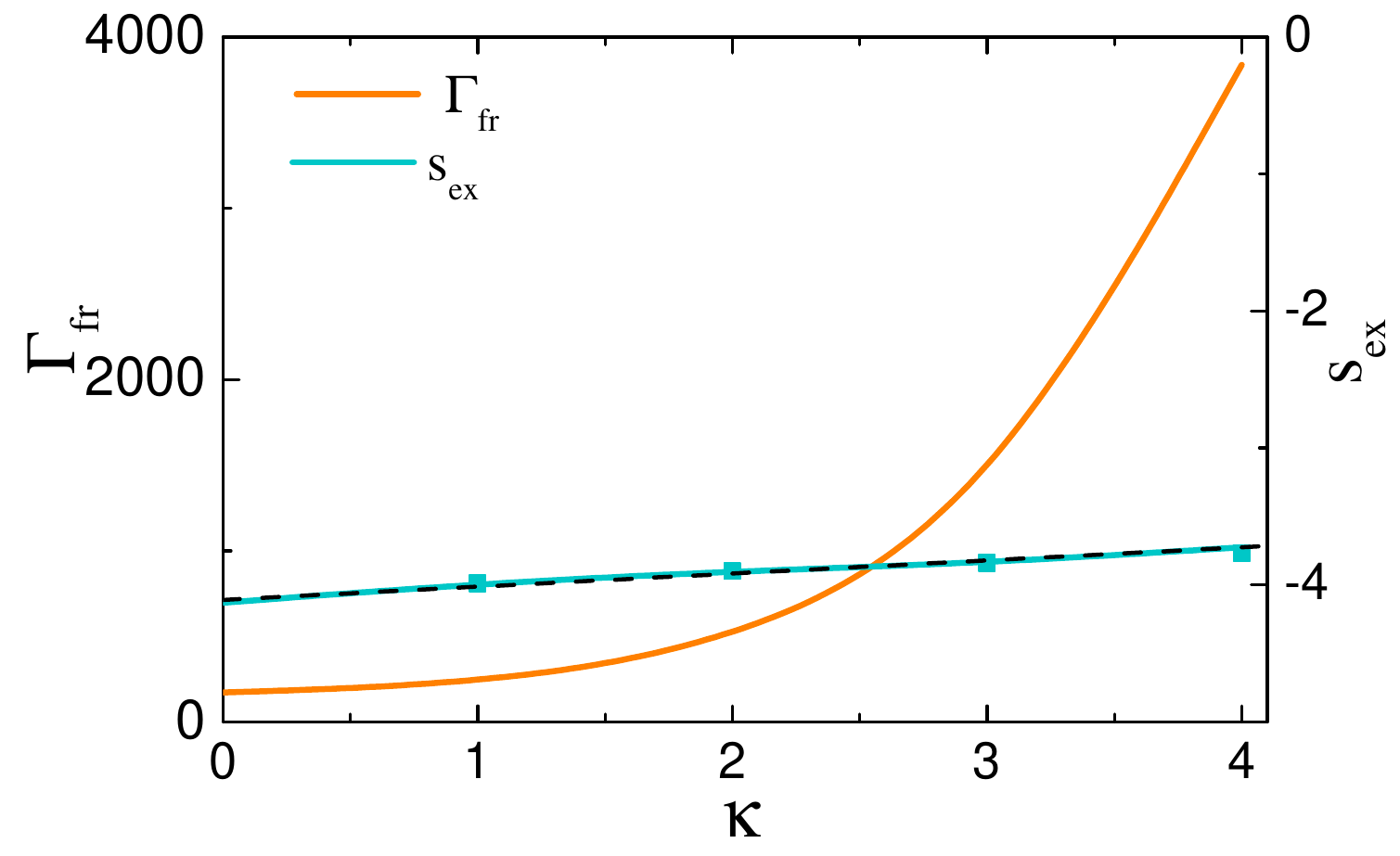}
\caption{(Color online) The coupling parameter (left axis) and the excess entropy (right axis) at freezing of the Yukawa fluid as functions of the screening parameter $\kappa$. The solid curves are plotted using the MD results of Ref.~\cite{HamaguchiPRE1997}. The squares are excess entropies evaluated from the vibrational model. The dashed line is a linear fit of the data points from Ref.~\cite{HamaguchiPRE1997}.}
\label{FigSFreezing}
\end{figure}

The coupling parameter and the excess entropy dependence on $\kappa$ at freezing of the Yukawa fluid are shown in Fig.~\ref{FigSFreezing}. Note that the excess entropy at freezing calculated with the help of he vibrational model is in excellent agreement with accurate results from numerical simulations~\cite{HamaguchiPRE1997}. While the freezing coupling parameter varies by more than 20 times as $\kappa$ increases from 0 to 4, the excess entropy varies only very weakly, systematically increasing with $\kappa$. This increase can be very well described by a linear function
\begin{equation}
s_{\rm ex}^{\rm fr}\simeq -4.109+0.096\kappa,
\end{equation} 
which is shown in Fig.~\ref{FigSFreezing} by the dashed line. This fit applies for $\kappa\leq 4$.

There is a simple instructive explanation why a quasi-universal value $s_{\rm ex}\simeq-4$ should be expected at the freezing point. Consider Eq.~(\ref{sex}) for the excess entropy within the vibrational model. Assuming for simplicity that all atoms vibrate with the same frequency (this is the Einstein approximation in the solid state physics) we get
\begin{equation}\label{sEinstein}
s_{\rm ex}\simeq\frac{3}{2}-\frac{3}{2}\ln \frac{m\Delta^2\Omega_{\rm E}^2}{2\pi T},
\end{equation}
where $\Omega_{\rm E}$ is the Einstein frequency. An average vibrational amplitude of an atom around its equilibrium position in a solid phase can be estimated within the framework of Einstein model via
\begin{equation}
\frac{1}{2}m\Omega_{\rm E}^2\langle \delta r^2\rangle = \frac{3}{2}T,
\end{equation}
which just reflects energy equipartition. The Einstein frequency in this harmonic approximation is thus related to the average vibrational amplitude as
\begin{equation}
\Omega_{\rm E}^2=\frac{3T}{m\langle \delta r^2\rangle}.
\end{equation}  
We can now estimate the Einstein frequency invoking the Lindemann's melting criterion arguments. The famous Lindemann’s criterion~\cite{Lindemann} states that a three-dimensional (3D) solid melts when the square root of the particle mean-squared displacement (MSD) from the equilibrium position reaches a threshold value, which is roughly $\sim 0.1$ of the interparticle distance. The Einstein frequency does not change much upon the fluid-solid phase transition. This is quite general statement, but for Yukawa systems (dusty plasma) it has been carefully verified experimentally~\cite{WongIEEE2018} and theoretically~\cite{KhrapakPoP03_2018} (although these works refer to 2D systems, the situation in 3D geometry is fully analogous). Substituting the Einstein frequency into Eq.~(\ref{sEinstein}) we get
\begin{equation}\label{ModifiedRT}
s_{\rm ex}\simeq \frac{3}{2}-\frac{3}{2}\ln\left( \frac{3}{2\pi}\frac{\Delta^2}{\langle \delta r^2\rangle}\right)\simeq -4.3, 
\end{equation}
where it is assumed that $\Delta^2/\langle \delta r^2\rangle\simeq 100$ at melting. This rough estimate is in very good agreement with the actual results presented in Fig.~\ref{FigSFreezing}. Our simple arguments are not limited to the Yukawa fluid, but apply equally well to other simple fluids~\cite{KhrapakPRR2020}, thus explaining quasi-universal values of the excess entropy at freezing.

\section{Modified Rosenfeld-Tarazona scaling}

Rosenfeld and Tarazona (RT) proposed a simple mathematical expression for the freezing temperature scaling of the thermal component of the excess internal energy of simple fluids~\cite{RosenfeldMolPhys1998}. Their derivation, based on the fundamental-measure free energy functional for hard spheres and thermodynamic perturbation theory, yields a quasi-universal high density expansion, featuring a
fluid Madelung energy with a $\propto T^{3/5}$ thermal correction. In reduced units their result can be presented as 
\begin{equation}
u_{\rm ex} = u_{\rm st}+u_{\rm th} \simeq M_{\rm fl}\Gamma+\delta\left(\frac{\Gamma}{\Gamma_{\rm fr}}\right)^{2/5},
\end{equation}    
where $u_{\rm st}$ is the static component of the excess energy, $u_{\rm th}$ is the thermal component, $M_{\rm fl}$ is the analogue of the Madelung energy for the fluid state and $\delta$ is a numerical coefficients. Rosenfeld further demonstrated that the numerical coefficients is also quasi-universal for different simple fluids and that the thermal correction can be represented as~\cite{RosenfeldPRE2000}
\begin{equation}\label{RT}
u_{\rm th}\simeq 3.0\left(\frac{\Gamma}{\Gamma_{\rm fr}}\right)^{2/5}.
\end{equation}  
This RT scaling, sometimes with minor modifications, has been successfully used to develop simple practical methods to calculate thermodynamic functions of Yukawa fluids as well as some related properties~\cite{KhrapakPRE02_2015,KhrapakPRE03_2015,KhrapakPPCF2015,KhrapakJCP2015,
ToliasPoP2015,ToliasPoP2019}. The thermal component of the excess energy, $u_{\rm th}$, gives direct access to the excess entropy by virtue of
\begin{equation}\label{sThermo}
\Gamma\frac{\partial s_{\rm ex}}{\partial \Gamma}=-u_{\rm th}+\Gamma\frac{\partial u_{\rm th}}{\partial \Gamma}.
\end{equation}
Combining equations (\ref{RT}) with (\ref{sThermo}) and integrating Rosenfeld obtained~\cite{RosenfeldPRE2000}
\begin{equation}\label{sRosenfeld}
s_{\rm ex}\simeq s_{\rm ex}^{\rm fr}-\frac{9}{2}\left[\left(\frac{\Gamma}{\Gamma_{\rm fr}}\right)^{2/5}-1\right],
\end{equation} 
where $s_{\rm ex}^{\rm fr}$ is the excess entropy at the freezing point. Equation (\ref{sRosenfeld}) delivers a correct value of $s_{\rm ex}$ at $\Gamma=\Gamma_{\rm fr}$, but does not generally reduces to $s_{\rm ex}=0$ at $\Gamma=0$ as desirable. As we discussed in Sec.~\ref{FigSFreezing}, the excess entropy at freezing of the Yukawa fluid is almost invariant, $s_{\rm ex}^{\rm fr}\simeq -4$. This means we will not make a big mistake by substituting the numerical coefficient $-\tfrac{9}{2}$ by $s_{ex}^{\rm fr}$ in Eq.~(\ref{sRosenfeld}). This modification of equation (\ref{sRosenfeld}) provides us with a reasonable interpolation between the weakly coupled and strongly coupled limits:
\begin{equation}\label{interpolation}
s_{\rm ex}\simeq s_{\rm ex}^{\rm fr}\left(\frac{\Gamma}{\Gamma_{\rm fr}}\right)^{2/5}.
\end{equation}

\begin{figure}
\includegraphics[width=8cm]{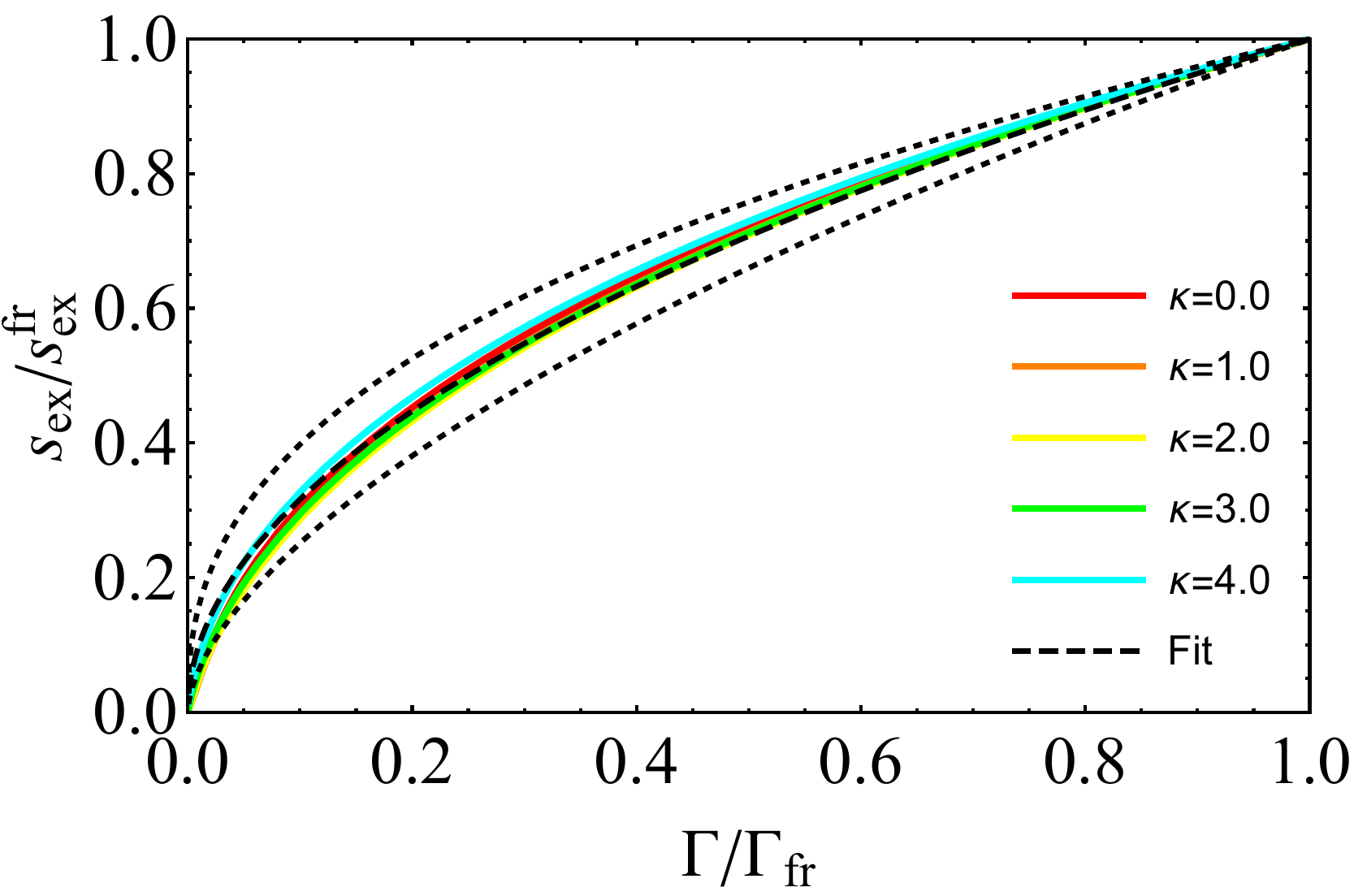}
\caption{(Color online) The ratio $s_{\rm ex}/s_{\rm ex}^{\rm fr}$ as a function of the reduced coupling strength $\Gamma/\Gamma_{\rm fr}$. The color curves correspond to the results from a MD simulation of Ref.~\cite{HamaguchiPRE1997}.  The dashed curve is the fit using equation (\ref{sfit}). The two dotted curves corresponding to the exponents $0.4$ and $0.6$ are plotted for comparison.}
\label{FigSex}
\end{figure}

Our important observation is that Eq.~(\ref{interpolation}) is {\it not the best option} to describe the actual dependence of the excess entropy on the reduced coupling strength. This is illustrated in Figure~\ref{FigSex}. Here the five color curves correspond to the excess entropy calculated using the fits of excess energy provided in Ref.~\cite{HamaguchiPRE1997} for fluids with $\kappa=0$ (one-component plasma limit), $\kappa =1$, $\kappa = 2$, $\kappa = 3$, and $\kappa = 4$. The first important observation is that the curves almost overlap each other, thereby confirming the universal character of the dependence of $s_{\rm ex}/s_{\rm ex}^{\rm fr}$ on $\Gamma/\Gamma_{\rm fr}$. The second observation is that a simple fit of the form     
\begin{equation}\label{sfit}
s_{\rm ex}\simeq s_{\rm ex}^{\rm fr}\left(\frac{\Gamma}{\Gamma_{\rm fr}}\right)^{1/2}
\end{equation}
describes excellently this universal dependence. We plot the two additional dotted curves corresponding to the dependence $s_{\rm ex}= s_{\rm ex}^{\rm fr}(\Gamma/\Gamma_{\rm fr})^{\gamma}$ with $\gamma = 0.4$ (the original RT scaling) and $\gamma=0.6$. This highlights the accuracy of the exponent $1/2$ in Eq.~(\ref{sfit}).    

What is essentially proposed is to use the scaling of $s_{\rm ex}$ with freezing temperature instead of $u_{\rm th}$ in the original formulation of the RT scaling. This simplifies some thermodynamic derivations. For instance, for the excess specific heat at constant volume we get    
\begin{equation}\label{cfit}
c_{\rm V}^{\rm ex}=-\Gamma\frac{\partial s_{\rm ex}}{\partial \Gamma}=-\frac{1}{2}s_{\rm ex}^{\rm fr}\left(\frac{\Gamma}{\Gamma_{\rm fr}}\right)^{1/2}.
\end{equation}
For the thermal component of excess energy we get in this approach
\begin{equation}
u_{\rm th}= \left(u_{\rm th}^{\rm fr}+s_{\rm ex}^{\rm fr}\right)\left(\frac{\Gamma}{\Gamma_{\rm fr}}\right)-s_{\rm ex}^{\rm fr}\left(\frac{\Gamma}{\Gamma_{\rm fr}}\right)^{1/2}.
\end{equation} 
This expression can be further simplified taking into account that for $s_{\rm ex}^{\rm fr}\simeq -4$ and $u_{\rm th}^{\rm fr}\simeq 3$. 

\section{Frenkel line on the phase diagram}

The existence and location of the Frenkel line, which marks the crossover between a gas-like and liquid-like dynamics is of considerable current interest (see e.g. Refs.~\cite{BrazhkinPRE2012,BrazhkinUFN2012,BrazhkinPRL2013,BrykJPCL2017,BellJCP2020,KhrapakJCP2022,
KhrapakPhysRep2024} and references therein). Various indicators of the crossover have been proposed over the years. Huang {\it et al}. performed extensive MD simulations of 3D and 2D Yukawa fluids and applied several diagnostics tools to locate the dynamical crossover~\cite{HuangPRR2023}. They demonstrated that the specific heat at constant volume $c_{\rm V}$ reaches the threshold value, oscillations in the velocity autocorrelation function emerge, the local atomic connectivity and the shear relaxation times become equal to the inverse Einstein frequency, the shear viscosity and the thermal conductivity coefficients have minima, all at the same relative coupling strength $\Gamma/\Gamma_{\rm fr}\simeq 0.05$. They also found that at the crossover the transverse (shear) sound velocity reaches a quasi-universal value of $c_t\simeq \sqrt{2T/m}$, which was then explained in Ref.~\cite{YuPRE2024}. 

The present author believes that one of the most convenient ways to distinguish between different dynamical regimes in simple fluids is by means of the excess entropy. A one-dimensional diagram of dynamical regimes sketched in Ref.~\cite{KhrapakPhysRep2024} features three distinct regimes.       
For $-1\lesssim s_{\rm ex} \lesssim 0$ the system is in gas-like state, and the gas–liquid dynamical crossover occurs at around $s_{\rm ex}= -1$. The region $-2\lesssim s_{\rm ex} \lesssim -1$ corresponds to the transitional regime where liquid-like properties start to prevail. For $s_{\rm ex} \lesssim -2 $ the vibrational paradigm of atomic dynamics applies. This corresponds to the dense fluid regime and this is where the vibrational model of transport coefficients and of excess entropy operates. The vibrational model is applicable up to the crystallization point, which roughly occurs at $s_{\rm ex} \simeq -4$, as discussed above. 

Adopting the value $\Gamma/\Gamma_{\rm fr}\simeq 0.05$ and using the scaling (\ref{sfit}) with $s_{\rm ex}^{\rm fr}\simeq -4$ we get $s_{\rm ex}\simeq -0.89$ at the location of the gas-liquid dynamical crossover in the Yukawa fluid. This is remarkably close to the value $s_{\rm ex}\simeq- 0.9\pm 0.1$ derived from the careful analysis of Stokes-Einstein relation in several simple fluids~\cite{KhrapakPRE10_2021,KhrapakJCP2022}. Using the scaling (\ref{cfit}) we also estimate $c_{\rm V}^{\rm ex}\simeq 0.45$ at the crossover, which is close to the threshold value $c_{\rm V}^{\rm ex}\simeq 0.5$~\cite{HuangPRR2023}.

\section{Excess entropy scaling of transport coefficients}

In 1977 Rosenfeld proposed a relation between transport coefficients and excess entropy of simple fluid systems~\cite{RosenfeldPRA1977}. In particular, he demonstrated that properly reduced diffusion and shear viscosity coefficients of several simple model fluids (hard sphere, soft sphere, one-component plasma, and Lennard-Jones fluids) are approximately exponential functions of the reduced excess entropy $s_{\rm ex}$. 
A somewhat different variant of entropy scaling of atomic diffusion in condensed matter was independently proposed by Dzugutov~\cite{DzugutovNature1996}. Later, Rosenfeld successfully applied his ideas of excess entropy scaling to transport coefficients of Yukawa fluids~\cite{RosenfeldPRE2000,RosenfeldJPCM2001}. At that time only a rather limited amount of transport data related to Yukawa fluids was available.  Many more numerical datasets on transport properties have been published in recent years. Taking into account the important role the Rosenfeld's excess entropy scaling plays in condensed matter nowadays~\cite{DyreJCP2018}, it makes sense to revisit its application to strongly coupled Yukawa fluids using modern transport data. 

The presented transport coefficients are normalized using the system-independent normalization adopted by Rosenfeld in his original work on the excess entropy scaling~\cite{RosenfeldPRA1977,RosenfeldJPCM1999}:
\begin{equation}\label{Rosenfeld}
D_{\rm R}  =  D\frac{\rho^{1/3}}{v_{\rm T}} , \quad
\eta_{\rm R}  =  \eta \frac{\rho^{-2/3}}{m v_{\rm T}}, \quad \lambda_{\rm R}  =  \lambda \frac{\rho^{-2/3}}{v_{\rm T}}, 
\end{equation}
where $v_{\rm T}=\sqrt{T/m}$ is the thermal velocity.

\begin{figure}
\includegraphics[width=8cm]{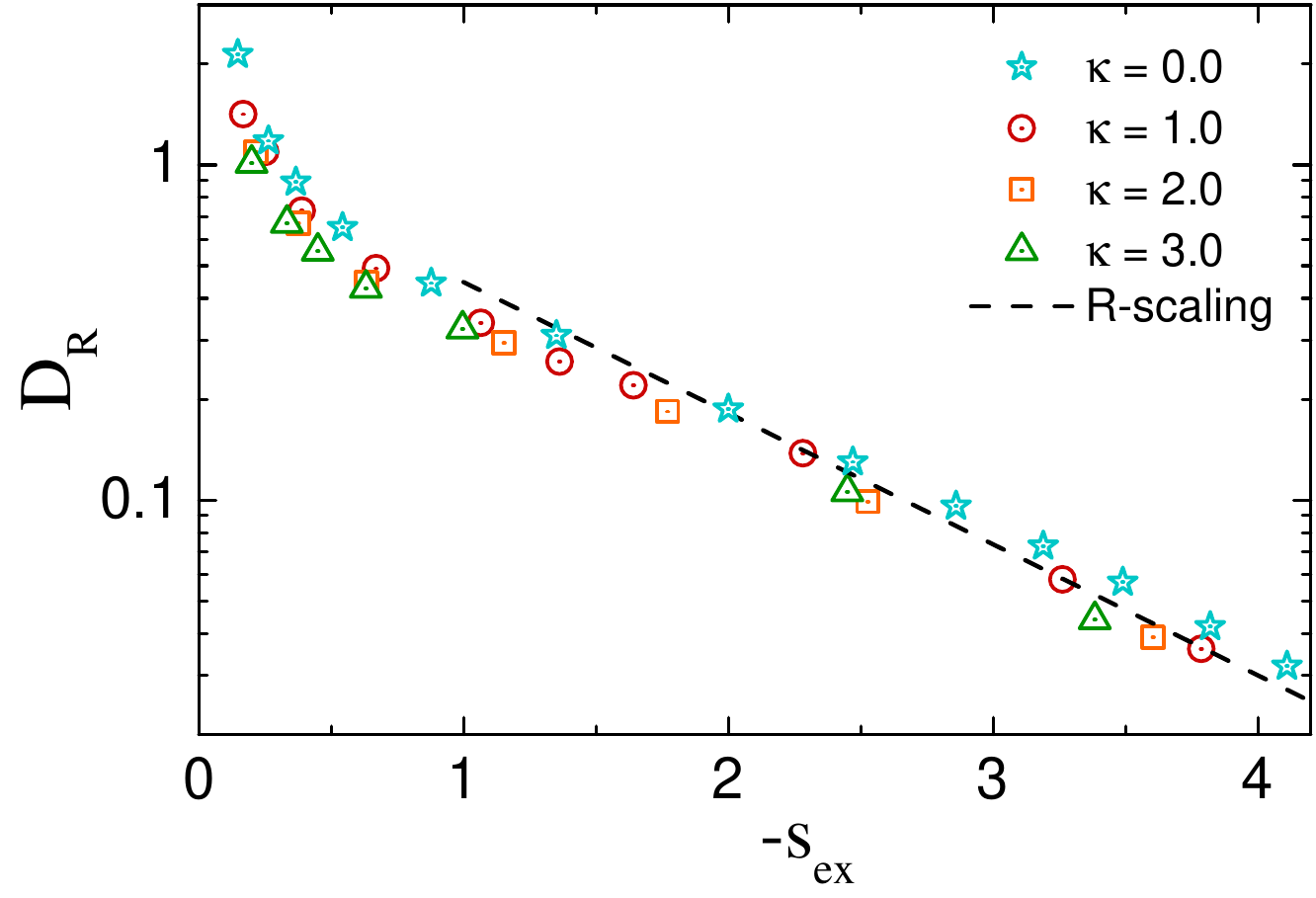}
\caption{(Color online) Reduced self-diffusion coefficient of the Yukawa fluid as a function of the negative excess entropy. Symbols correspond to the numerical results from Ref.~\cite{OhtaPoP2000}. The dashed line is an exponential fit of Eq.~(\ref{Dfit}).}
\label{FigDiff}
\end{figure}

Figure~\ref{FigDiff} presents the results corresponding to the self-diffusion coefficient $D_{\rm R}$ of the Yukawa fluid. The symbols are the numerical simulation results from Ref.~\cite{OhtaPoP2000}. They are nicely collapsing to a quasi-universal master curve. In the strongly coupled regime, the self-diffusion coefficient can be reasonably well described by an exponential function 
\begin{equation}\label{Dfit}
D_{\rm R}\simeq 1.1\exp\left(0.9 s_{\rm ex}\right).
\end{equation}
Note that the onset of validity of this fit, $s_{\rm ex}\lesssim -1$, corresponds to the crossover between the gas-like and liquid-like dynamics (Frenkel line).  

\begin{figure}
\includegraphics[width=8cm]{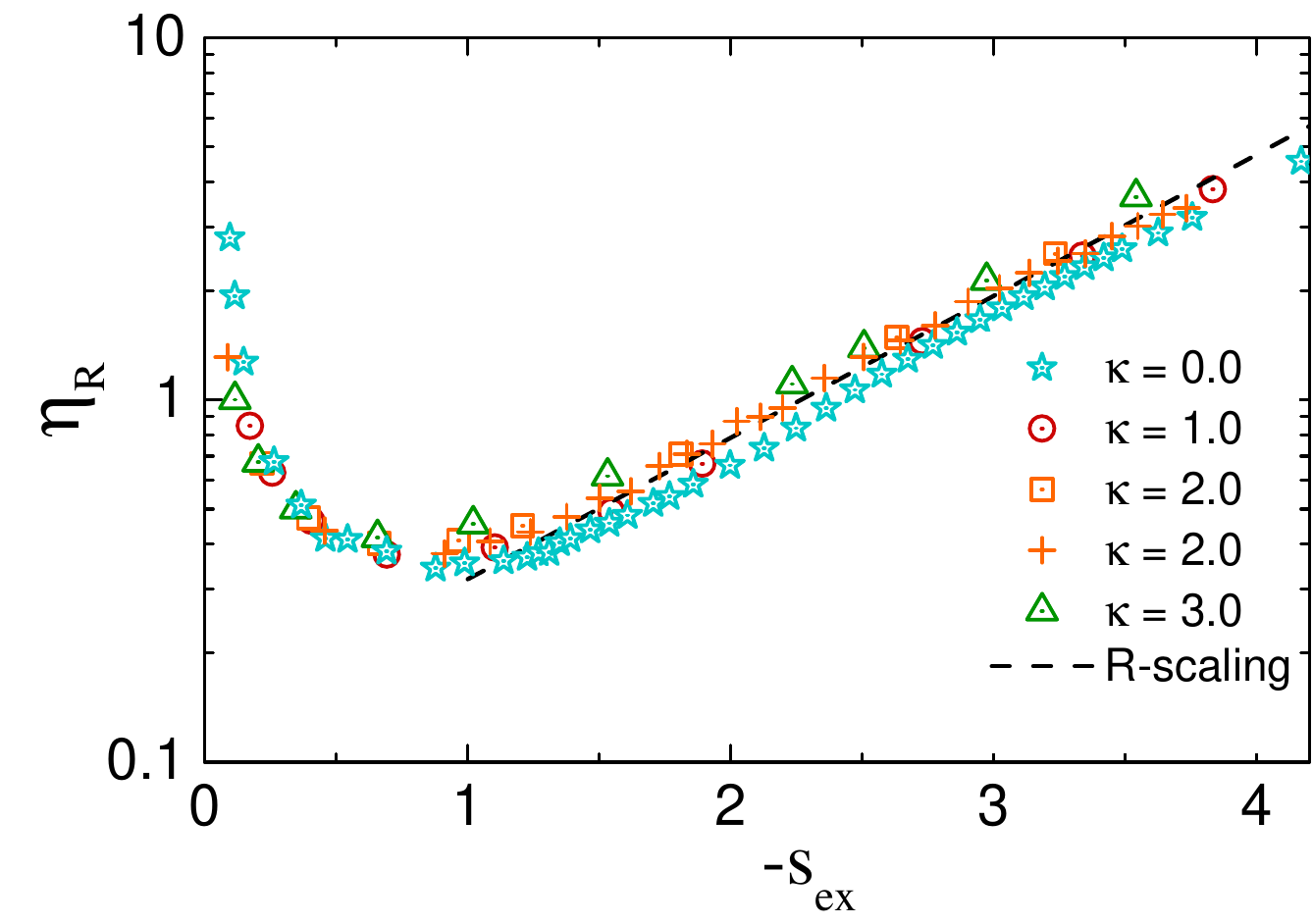}
\caption{(Color online) Reduced shear viscosity coefficient of the Yukawa fluid as a function of the negative excess entropy. Symbols correspond to the numerical results from Ref.~\cite{DonkoPRE2008,DaligaultPRE2014}. The dashed line is an exponential fit of Eq.~(\ref{Etafit}).}
\label{FigEta}
\end{figure}

The shear viscosity ($\eta_{\rm R}$) data are summarized in Fig.~\ref{FigEta}. The symbols correspond to the numerical results from Refs.~\cite{DonkoPRE2008,DaligaultPRE2014}. Similar to the case of self-diffusion, they nicely collapse to a quasi-universal curve. In the strongly coupled regime, the reduced shear viscosity coefficient can be reasonably well fitted by an exponential function
\begin{equation}\label{Etafit}
\eta_{\rm R}\simeq 0.13\exp\left(-0.9 s_{\rm ex}\right).
\end{equation}
By construction, equations (\ref{Dfit}) and (\ref{Etafit}) satisfy the Stokes-Einstein relation without the hydrodynamic diameter~\cite{CostigliolaJCP2019},
\begin{equation}
D_{\rm R}\eta_{\rm R}\simeq 0.14,
\end{equation} 
as they should~\cite{KhrapakMolPhys2019,KhrapakPRE10_2021}. The onset of this exponential dependence is again at the gas-liquid dynamical crossover, at $s_{\rm ex}\simeq -1$. This is roughly where the minimum in $\eta_{\rm R}$ is reached~\cite{KhrapakPoF2022}. 

Note that substituting an approximate scaling $s_{\rm ex}\simeq -4(\Gamma/\Gamma_{\rm fr})^{1/2}$ into Eq.~(\ref{Etafit}) we get $\eta_{\rm R}\simeq 0.13\exp[3.6(\Gamma/\Gamma_{\rm fr})^{1/2}]$. This essentially coincides with an empirical fit for the reduced viscosity coefficient of the strongly coupled Yukawa fluid proposed in Ref.~\cite{KhrapakAIPAdv2018}, based on the ideas from Ref.~\cite{CostigliolaJCP2018}. Now it becomes evident that this empirical fit is in fact consistent with the Rosenfeld's version of excess entropy scaling.   

\begin{figure}
\includegraphics[width=8cm]{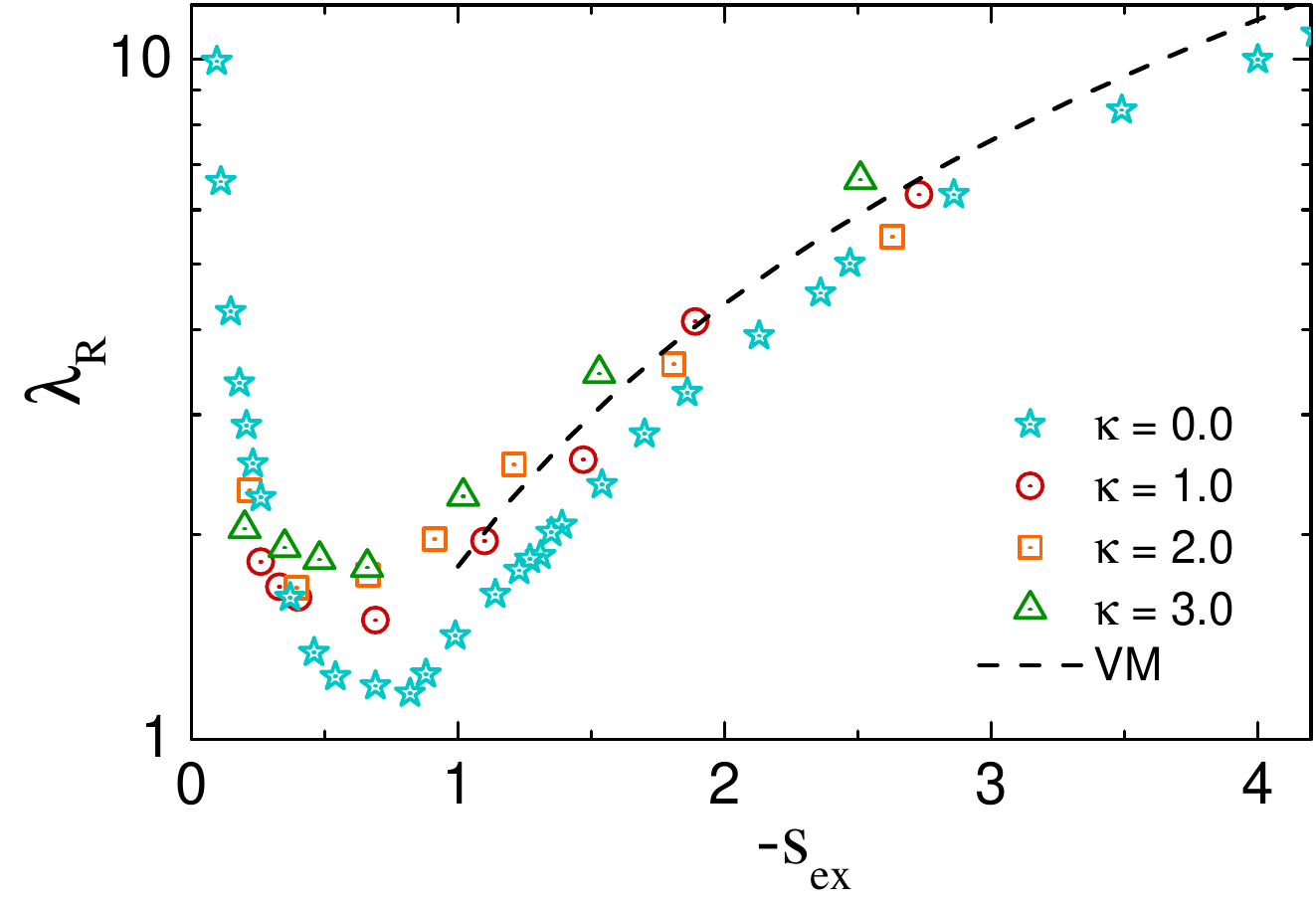}
\caption{(Color online) Reduced thermal conductivity coefficient of the Yukawa fluid as a function of the negative excess entropy. Symbols correspond to the numerical results from Ref.~\cite{DonkoPRE2004,ScheinerPRE2019}. The dashed line corresponds to the vibrational model of heat transfer, Eq.~(13) from Ref.~\cite{KhrapakPPR2023}, under the assumption $s_{\rm ex}\simeq -4(\Gamma/\Gamma_{\rm fr})^{1/2}$.}
\label{FigLambda}
\end{figure}

The thermal conductivity ($\lambda_{\rm R}$) data are summarized in Fig.~\ref{FigLambda}. The symbols correspond to the numerical results from Refs.~\cite{DonkoPRE2004,ScheinerPRE2019}. In this case the collapse of the data points is not so impressive as for the diffusion and viscosity coefficients. The reason is unclear at this point. In other fluids, such as for instance the Lennard-Jones fluid, the quality of collapse is apparently better~\cite{BellJPCB2019} (if near-critical effects are not considered). More data on the thermal conductivity coefficient of the Yukawa fluid would be most welcome and will help to understand the quality of the excess entropy scaling in this case. The dashed curve in Fig.~\ref{FigLambda} corresponds to the  estimation using the vibrational model of heat transfer~\cite{KhrapakPoP08_2021,KhrapakPPR2023} assuming $s_{\rm ex}\simeq -4(\Gamma/\Gamma_{\rm fr})^{1/2}$. The agreement with the numerical results is fair. The minimum in $\lambda_{\rm R}$ is reached relatively close to $s_{\rm ex}\simeq -1$, where the gas-liquid dynamical crossover takes place.

\section{Conclusion}

The most important results of this study can be formulated as follows.

The vibrational model of atomic dynamics in dense fluids appears quite successful to describe the excess entropy of the screened Coulomb (Yukawa) fluid. Without any adjustable parameters the model allows to estimate the excess entropy with a relatively high accuracy. The model provides a direct link between collective properties and thermodynamics and as such contributes to better understanding of the liquid state.  

The excess entropy at freezing of Yukawa fluids is almost constant, $s_{\rm ex}\simeq -4$. This property apparently applies to many other simple fluids and there is a solid explanation why this     
should be approximately so. A weak dependence on the screening parameter is nevertheless still present and can be accounted for by a simple linear function.

A practical scaling of the excess entropy with the freezing temperature is put forward, which can be considered as the modification of the Rosenfeld-Tarazona scaling. It is observed that for the Yukawa fluid, the excess entropy scales as $s_{\rm ex}\simeq s_{\rm ex}^{\rm fr}(\Gamma/\Gamma_{\rm fr})^{1/2}$, instead of the original RT exponent equal to $2/5$. Combination of the weak linear dependence of $s_{\rm ex}^{\rm fr}$ on $\kappa$ with the modified Rosenfeld-Tarazona scaling represents a simple practical tool to estimate the excess entropy across the the entire phase diagram of Yukawa fluids.

The location of the gas-to-liquid dynamical crossover in the Yukawa fluid is discussed. The excess entropy criterion $s_{\rm ex}\simeq -1$ agrees well with several other measures proposed in the literature.  

Excess entropy scaling of transport coefficients works rather well for the self-diffusion and shear viscosity coefficients. The approximate formulas based on modern transport datasets are suggested. The Stokes-Einstein relation  without the hydrodynamic radius between the self-diffusion and shear viscosity coefficients is satisfied at strong coupling, as expected. The scaling of the thermal conductivity coefficient with excess entropy is not so impressive. More data is needed to resolve this issue and this might be a reasonable suggestion for future simulations.     

The ideas presented in this work can be straightforwardly extended to bi-
Yukawa systems (relevant for complex plasmas) and to any multi-Yukawa systems, at least
when all interaction terms are purely repulsive. This is indicative of the generality of the
results. In fact, the RT decomposition and scaling have already been utilized
for the construction of a very accurate bi-Yukawa equation of state for the excess internal energy~\cite{CastelloPoP2019}.

% (in spite of the involvement of two external parameters and two state variables in the pair
%interaction potential) 

Overall, the present results provide a systematic picture on the useful interrelations between the properties of collective modes, thermodynamics and transport in simple fluids with soft pairwise repulsive interactions. They can be of particular interest in the context of complex (dusty) plasma, colloidal suspensions, electrolytes, and other related soft matter systems.

\appendix*

\section{Reference data for the excess entropy}

In Ref.~\cite{HamaguchiPRE1997} the excess entropy data is not directly available. What is presented is the equilibrium excess internal energy at $\Gamma\geq 1$, including the interaction with the neutralizing background, in the form of fitting expressions
\begin{equation}
u_{\rm ex}(\kappa, \Gamma)= a(\kappa)\Gamma + b(\kappa)\Gamma^{1/3}+c(\kappa)+d(\kappa)\Gamma^{-1/3}.
\end{equation} 
The coefficients $a$, $b$, $c$, and $d$ are tabulated for several values of $\kappa$ in the range $1.2\leq \kappa \leq 5$ in Tab. VIII of Ref.~\cite{HamaguchiPRE1997}. The regime $\kappa\leq 1.0$ is treated separately in Ref.~\cite{FaroukiJCP1994}. From the excess energy we can calculate the excess Helmholtz free energy by means of the integration
\begin{equation}
f_{\rm ex}(\kappa,\Gamma)=\int_1^{\Gamma}u_{\rm ex}(\kappa,\Gamma')\frac{d\Gamma'}{\Gamma'}+f_1(\kappa),
\end{equation}  
where
\begin{equation}
f_1(\kappa)=\int_0^{1}u_{\rm ex}(\kappa,\Gamma')\frac{d\Gamma'}{\Gamma'}
\end{equation} 
is obtained using the actual dependence of $u_{\rm ex}(\kappa,\Gamma)$ on $\Gamma$ in the weakly coupled regime $\Gamma\leq 1$. The numerical values of $f_1(\kappa)$ are listed in Tab. VII of Ref.~\cite{HamaguchiPRE1997} for $0.0\leq \kappa\leq 5.0$. The reduced excess entropy is then obtained from
\begin{equation}
s_{\rm ex}(\kappa,\Gamma) = u_{\rm ex}(\kappa,\Gamma)-f_{\rm ex}(\kappa,\Gamma).
\end{equation} 

%\acknowledgments

%\bibliographystyle{aipnum4-1}

\bibliography{SE_Ref}

\end{document}